\documentclass[pra,
twocolumn,
floatfix]{revtex4-2}
\usepackage[utf8]{inputenc}
\usepackage{color}

\usepackage{lmodern}
\usepackage{amsmath}
\usepackage{bm}
\usepackage{amssymb}
\usepackage{mathtools}
\usepackage{hyperref}

\usepackage[english]{babel} 

\usepackage[tight,nice]{units}

\usepackage{pgfplots}
\pgfplotsset{compat=newest}
\usepgfplotslibrary{groupplots}
\usetikzlibrary{arrows.meta}
\usetikzlibrary{positioning}

\DeclareMathAlphabet\mathbfcal{OMS}{cmsy}{b}{n}

\newcommand{\Real}{\mathrm{Re}}
\newcommand{\Imag}{\mathrm{Im}}

\DeclareMathAlphabet{\bi}{OML}{cmm}{b}{it}
\renewcommand{\vec}[1]{ \bi{ #1 } }

\newcommand{\rmi}{\mathrm{i}}

\newcommand{\imagi}{\rmi}

\newcommand{\diff}{\mathrm{d}}

\newcommand{\abl}[2]{\frac{\diff #1}{\diff #2}}

\newcommand{\twovec}[2]{\left( \begin{array}{c} #1 \\ #2 \end{array}\right)}

\newcommand{\eulere}{\mathrm{e}}

\newcommand{\beq}{\begin{equation}}
\newcommand{\eeq}{\end{equation}}

\newcommand{\ket}[1]{\vert #1\rangle}
\newcommand{\bra}[1]{\langle#1\vert}
\newcommand{\braket}[2]{\langle #1 \vert #2 \rangle}

\renewcommand{\Re}{\,\mathrm{Re}\,}
\renewcommand{\Im}{\,\mathrm{Im}\,}
\DeclareMathOperator\Arg{arg}

\newcommand{\threevec}[3]{\left( \begin{array}{c} #1 \\ #2 \\ #3 \end{array}\right)}

\def\twobytwo#1#2#3#4{\left( \begin{array}{cc} #1 & #2 \\ #3 & #4 \end{array}\right)}

\begin{document}

\title{Intense-laser driven electron dynamics and high-harmonic generation in solids including topological effects}

\author{Daniel Moos}
\author{Christoph J\"ur{\ss}}
\author{Dieter Bauer}
\affiliation{Institut f\"ur Physik, Universit\"at Rostock, 18051 Rostock, Germany}

\date{\today}

\begin{abstract} 
A theory for laser-driven electron dynamics and high-harmonic generation in bulk solids with two lattice sites per unit cell of arbitrary dimension is formulated. In tight-binding approximation, such solids can be described by $2\times 2$ Bloch-Hamiltonians. Our theory is able to fully capture topological effects in high-harmonic generation by such systems because no simplifications beyond tight-binding, dipole approximation, and negligible depletion of the valence band are made. An explicit, analytical expression for the electron velocity 
is given. Exemplarily, the theory is applied to the Su-Schrieffer-Heeger chain and the Haldane model in strong laser fields. 
\end{abstract}

\maketitle

\section{Introduction}
Topological phase transitions are intimately related to the closing and reopening of band gaps \cite{Witten2016,topinsshortcourse,topinsRevModPhys.82.3045} as a function of parameters in the Hamiltonian. Above and below such a phase transition, the electron dynamics of the system may qualitatively change drastically, e.g., from clockwise to counter-clockwise motion. It is not surprising that such changes affect the radiation emitted by the electrons.

 The light emitted by intense-laser-driven electrons may contain high harmonics of the incident laser pulse's carrier frequency. High-harmonic generation (HHG) by isolated atoms or molecules in the gas phase has been extensively studied over the last decades and is the basis for the synthesis of attosecond pulses, which then can be used to explore ultrafast dynamics directly in the time domain \cite{attoKrausz}. HHG in solids was observed at modest laser intensities below the destruction threshold \cite{Ghimire2011,ndabashimiye_solid-state_2016}. It was soon shown that the harmonics contain structural information and thus allows for an all-optical probing of condensed matter \cite{VampaPhysRevLett.115.193603,LangerF.2017}, including the measurement of Berry curvature \cite{PhysRevB.96.075409,Luu2018}. Very recently, it has been shown that the valence electron density can be probed by HHG \cite{Lakhotia2020}, complementing the conventional crystallographic methods that probe the ion positions. 

A wide area of modern condensed matter physics is concerned with geometrical phases and topology \cite{topins,topinsbernevig,vanderbiltbook}. Hence, from the strong-field, attosecond perspective, the natural question arises whether topological effects can be probed using, e.g., HHG, or exploited to affect the strong-field electron dynamics. While predictions are difficult, the ultrafast steering of currents by lasers \cite{Schultze2013,Garg2016,Higuchi2017,Baudisch2018}, in particular topologically protected edge currents \cite{Reimann2018}, the modification of topological properties via laser dressing \cite{Hubener2017,giovannini2020}, and ultrafast valleytronics \cite{jimnezgaln2019lightwave} are probably the most promising mergers of modern condensed matter physics and strong-field attosecond science so far.

As topological phase transitions are related to the closing and reopening of band gaps, the simplest but nontrivial systems have two bands. It is further known that topologically nontrivial phases in the bulk lead to edge states in the corresponding finite system (``bulk-boundary correspondence'' \cite{topinsshortcourse,vanderbiltbook}). However, HHG in finite solids with explicit edge states, while definitely very interesting and with huge topological effects in HHG found numerically \cite{bauer_high-harmonic_2018,DrueekeRobustness2019,JuerssSSH2019},
 is hardly accessible analytically. Hence, in this work, we concentrate on the bulk so that a Bloch ansatz can be made, reducing the problem to a $2\times 2$ Bloch-Hamiltonian. Prominent examples covered by this approach are, e.g., the Su-Schrieffer-Heeger (SSH) chain \cite{SSHPhysRevLett.42.1698,topinsshortcourse}, graphene \cite{RevModPhys.81.109}, the Haldane \cite{Haldane88} or the Qi-Wu-Zhang \cite{QiWuZhang06,topinsshortcourse} model. While the electronic structure of these systems and their topological properties are well studied, the investigation of ultrashort, strong-field electron dynamics and HHG in them have started only recently \cite{PhysRevB.96.075409,bauer_high-harmonic_2018,DrueekeRobustness2019,JuerssSSH2019,Luu2018,chacon_observing_2018,Silva2019,jimnezgaln2019lightwave}.

In this paper, we aim at providing the theoretical minimum of laser-driven electron dynamics and HHG in solids. Besides restricting ourselves to two bands neither assumptions about, e.g., the dimensionality, are made nor do we approximate transition matrix elements because this may sweep topological effects under the rug. The main result of this paper is an explicit expression for the laser-driven electron velocity as a function of the system-specific three-vector $\vec d(\vec k)$ (see equation \eqref{eq:2x2BlochHamil} below) and the driving laser field. The HHG spectrum can then be calculated from the Fourier-transform of the velocity, acceleration, or current \cite{bandrauk_quantum_2009,baggesen_dipole_2011,bauer_computational_2017}.

The outline of the paper is as follows. In Section \ref{sec:theory}, we introduce our theory, including quick reminders about tight-binding, Bloch-Hamiltonians, and the coupling of tight-binding Hamiltonians to laser fields. In Section \ref{sec:couplingandEOMs}, we also derive the equations of motion to be solved for the calculation of HHG spectra. The electron velocity is calculated in Section \ref{sec::velocity} before, in Section \ref{sec:lewensteinmodelling}, the analogies to gas-phase HHG are briefly discussed. In Sections \ref{sec:resultsSSH} and \ref{sec:resultsHaldane}, we use our theory to calculate HHG spectra for the SSH and Haldane model, respectively. The purpose of these results is twofold. First, we had to check that our main result, i.e., the analytical expression for the laser-driven electron velocity, is correct by comparison with the numerical solutions of the equations of motion in position or $\vec k$ space. Second, we want to trigger more interest in ``topological HHG'' by illustrating the counter-intuitive electron motion in condensed matter.
We conclude in Section \ref{sec:summ} and give some details on the proper choice of the Bloch ansatz and the Haldane model in the Appendix.

\section{Theory} \label{sec:theory}
In the following subsections we introduce the theory underlying our calculations of high-harmonic spectra from solids and set the stage notation wise. Atomic units $\hbar=|e|=m_e=1$ are used unless indicated otherwise.

\subsection{Tight-binding}
Starting from a continuous description of a solid, a lattice Hamiltonian is obtained by a tight-binding ansatz \cite{vanderbiltbook}
\beq \phi_{\vec R\alpha}(\vec r) = \varphi_\alpha(\vec r-\vec R - \boldsymbol{\tau}_\alpha), \qquad \alpha=1,2,\ldots, M . \label{eq:tb}\eeq
Here, $\vec R=\vec R_{n_1 \dots n_D} =\sum_j n_j\vec a_j$ is a lattice vector pointing to some unit cell defined by the basis vectors $\vec a_j$, $j=1,2, \ldots, D$, of the $D$-dimensional system, the index $\alpha$ labels the $M$ orbitals $\varphi_\alpha$ per unit cell, with $\boldsymbol{\tau}_\alpha$ fixing their position within the unit cell. We assume
\beq \braket{\phi_{\vec R\alpha}}{\phi_{\vec R'\beta}} = \delta_{\vec R\vec R'} \delta_{\alpha\beta} , \eeq
i.e., the orbitals should be orthonormal both within a unit cell and across unit cells. Making use of the discrete translational invariance of $\hat H$ one can write the Hamiltonian $\hat H = \sum_{\vec R\alpha} \ket{\phi_{\vec R\alpha}}\bra{\phi_{\vec R\alpha}} \hat H  \sum_{\vec R'\beta}\ket{\phi_{\vec R'\beta}}\bra{\phi_{\vec R'\beta}}$ in ``hopping form'' 
\beq \hat H = \sum_{\Delta\vec R}\sum_{\alpha\beta} H_{\alpha\beta}(\Delta\vec R) \sum_{\vec R} \ket{\phi_{\vec R\alpha}} \bra{\phi_{\Delta\vec R+\vec R,\beta}} , \label{eq:tbH}\eeq
where $\Delta\vec R = \vec R'-\vec R$ and $H_{\alpha\beta}(\Delta\vec R)=\bra{\phi_{\vec R\alpha}} \hat H  \ket{\phi_{\vec R'\beta}}$. The Hamiltonian \eqref{eq:tbH} is a sum over hoppings from cell $\vec R + \Delta \vec R$ and orbital $\beta$ to cell $\vec R$ and orbital $\alpha$, weighted by the matrix element $H_{\alpha\beta}(\Delta\vec R)$.

With the Bloch ansatz
\beq \ket{\phi_{\vec k \alpha}} = \sum_{\vec R} \eulere^{\imagi\vec k\cdot(\vec R + \boldsymbol{\tau}_\alpha)} \ket{\phi_{\vec R\alpha}}, \label{eq:Blochlikeansatz} \eeq
i.e.,
\beq  \ket{\phi_{\vec R\alpha}} = \frac{V_{\mathrm{cell}}}{(2\pi)^D} \int_{\mathrm{BZ}}\diff^D k\, \eulere^{-\imagi\vec k\cdot(\vec R+ \boldsymbol{\tau}_\alpha)}\ket{\phi_{\vec k \alpha}}  \eeq
and
\beq \braket{\phi_{\vec k \alpha}}{\phi_{\vec k' \beta}} = \frac{(2\pi)^D}{V_{\mathrm{cell}}} \delta^D(\vec k-\vec k') \delta_{\alpha\beta}, \label{eq:kalphakbetaortho}\eeq
where $\int_{\mathrm{BZ}}\diff^D k$ is the integral over the Brillouin zone and $V_{\mathrm{cell}}$ is the volume of the $D$-dimensional unit cell,
follows
\beq\hat H =  \frac{V_{\mathrm{cell}}}{(2\pi)^D}\int_{\mathrm{BZ}}\diff^D k\sum_{\alpha\beta} H_{\alpha\beta}(\vec k) \, \ket{\phi_{\vec k \alpha}}\bra{\phi_{\vec k \beta}}, \label{eq:Hhatinkspace}
\eeq
where
\beq H_{\alpha\beta}(\vec k) = \sum_{\Delta\vec R} H_{\alpha\beta}(\Delta\vec R)\, \eulere^{\imagi\vec k\cdot (\Delta\vec R+\boldsymbol{\tau}_\beta - \boldsymbol{\tau}_\alpha)  } \label{eq:BlochHamilTB} \eeq
is the tight-binding Bloch-Hamiltonian.

The eigenvalues $E_n(\vec k)$ of $H_{\alpha\beta}(\vec k)$ for all $\vec k$ within the first Brillouin zone will give the band structure consisting, in general, of $n=1,2, \ldots, M$ bands. We can expand the eigenstates as
\beq \ket{\psi_{n\vec k}} = \sum_\alpha C^\alpha_n(\vec k) \ket{\phi_{\vec k \alpha}}, \label{eq:psink}\eeq
where $n$ is the band index. We normalize $\sum_\alpha |C_n^\alpha(\vec k)|^2=1$ so that
\beq \braket{\psi_{n\vec k'}}{\psi_{n\vec k}}= \frac{(2\pi)^D}{V_{\mathrm{cell}}} \delta^D(\vec k -\vec k') . \eeq
Plugging this into the time-independent Schr\"odinger equation \beq E_n(\vec k) \ket{\psi_{n\vec k}} = \hat H \ket{\psi_{n\vec k}} \label{eq:anotherSE}\eeq with the Hamiltonian \eqref{eq:Hhatinkspace}, using equation \eqref{eq:kalphakbetaortho}, and
multiplication from the left by $\bra{\phi_{\vec k \alpha}}$ leads to
\beq E_n(\vec k)C^\alpha_n(\vec k) = \sum_{\gamma} H_{\alpha\gamma}(\vec k) \, C^\gamma_n(\vec k) \eeq
or, in matrix notation,
\beq E_n(\vec k) \mathbf{C}_n(\vec k) = \mathbf{H} (\vec k) \mathbf{C}_n(\vec k) . \label{eq:matrixBloch} \eeq

\subsection{Case of a $2\times 2$ Bloch-Hamiltonian}
If there are only two orbitals $\alpha=1,2$ per unit cell in \eqref{eq:tb}, may they be the two ground states of two atoms at different positions $\boldsymbol{\tau}_\alpha$ or the lowest two states in one atom per unit cell, the Bloch-Hamiltonian matrix $\mathbf{H} (\vec k)$ in \eqref{eq:matrixBloch} is $2\times 2$, leading to two bands. Topological effects arise because of the closing and reopening of a band gap as a function of some parameter in $\mathbf{H} (\vec k)$. Hence, the study of the two bands whose band gap closes and reopens and the corresponding $2\times 2$ Bloch-Hamiltonian is usually sufficient. Prime examples for such systems described by $2\times 2$ Bloch-Hamiltonians are the SSH chain \cite{SSHPhysRevLett.42.1698,topinsshortcourse} and the Haldane model \cite{Haldane88} both of which will be discussed in this work in the context of HHG.

Writing $n=\pm$ instead of $n=1,2$ for the two bands, where $+$ denotes the energetically higher conduction band and $-$ the lower valence band, the eigenvalue equation \eqref{eq:matrixBloch} becomes
\beq  E_{\pm}(\vec k) \mathbf{C}_\pm(\vec k) = \mathbf{H}(\vec k) \mathbf{C}_\pm(\vec k) \label{eq:BlochforSSHrepeat}.\eeq
The hermitian $2\times 2$ Bloch-Hamiltonian can be expanded in Pauli matrices,
\beq \mathbf{H}(\vec k) = \vec d(\vec k) \cdot \boldsymbol{\sigma}  \label{eq:2x2BlochHamil} \eeq
where $\vec d(\vec k)=(d_x(\vec k),d_y(\vec k),d_z(\vec k))^\top\in\mathbb{R}^3$ is a three-component vector, and $\boldsymbol{\sigma}=(\boldsymbol{\sigma}_x,\boldsymbol{\sigma}_y,\boldsymbol{\sigma}_z)^\top$ is the three-component vector of Pauli matrices
\beq \boldsymbol{\sigma}_x=\twobytwo{0}{1}{1}{0},\quad  \boldsymbol{\sigma}_y=\twobytwo{0}{-\imagi}{\imagi}{0},  \quad  \boldsymbol{\sigma}_z=\twobytwo{1}{0}{0}{-1} . \eeq
There could be also a term $d_0 \boldsymbol{1}$ proportional to the $2\times 2$ unity matrix. However, in this work we only discuss systems with $d_0 = 0$.
Further, we consider only hermitian $\mathbf{H}(\vec k)$ in this work so that $\vec d(\vec k)$ is real.

The tight-binding representation of the system in position space might be of arbitrary dimension $D$, resulting in $D$ components of the lattice momentum $\vec k$. For the SSH model, $\vec k$ has just one component $k$, and for the Haldane model, $\vec k\in\mathbb{R}^2$. Anyhow, we can write
\beq  \mathbf{H}(\vec k) =  \twobytwo{d_z(\vec k)}{d_x(\vec k) - \imagi d_y(\vec k)}{d_x(\vec k)+\imagi d_y(\vec k)}{-d_z(\vec k)} \label{eq:general2x2}
 \eeq
with eigenvalues
\beq E_\pm(\vec k) = \pm |\vec d(\vec k)| = \pm d(\vec k) . \eeq
We recognize that the specific information about the actual system under consideration lies in the dependence of $\vec d$ on $\vec k$. As long as we are general, we will suppress the $\vec k$-dependence in the expressions, i.e.,
 \beq  \mathbf{H}(\vec d) =  \twobytwo{d_z}{d_x - \imagi d_y}{d_x+\imagi d_y}{-d_z} \label{eq:general2x2asfuncofd}
 \eeq
with eigenvalues
\beq E_\pm(\vec d) = \pm d . \eeq
We assume the eigenvectors $\mathbf{C}_\pm(\vec k)$ to be normalized, which still leaves the structure-gauge freedom \cite{YueGaarde2020} that allows for transformations \beq \mathbf{C}'_\pm(\vec k) = \exp[\imagi\chi_\pm(\vec k)]\mathbf{C}_\pm(\vec k), \label{eq:structuregaugetrafo} \eeq with some real, differentiable functions $\chi_\pm(\vec k)$, without affecting observables such as velocity or current.
A possible eigenvector belonging to $E_+(\vec d)$ is
\beq \mathbf{C}_+(\vec d) = \frac{1}{\sqrt{2}}\twovec{\sqrt{1+d_z/d}}{\frac{d_x+ \imagi d_y}{d_\perp}\sqrt{1-d_z/d}} , \label{eq:generalpossibleCplus}\eeq
and for $E_-(\vec d)$
\beq  \mathbf{C}_-(\vec d)  = \mathbf{C}_+(-\vec d)= \frac{1}{\sqrt{2}}\twovec{\sqrt{1-d_z/d}}{-\frac{d_x+ \imagi d_y}{d_\perp}\sqrt{1+d_z/d}} \label{eq:generalpossibleCminus} \eeq
where $d_\perp=\sqrt{d_x^2+d_y^2}$.
In the literature, the topological properties of the $2\times 2$ Bloch-Hamiltonian are conveniently discussed using the Bloch-sphere angles $\theta,\varphi$, i.e., $\cos\theta = {d_z}/{d}$ and 
$\eulere^{\imagi\varphi} = {(d_x+\imagi d_y)}/({d \sin\theta })$ \cite{topinsshortcourse,topinsbernevig}. However, there is no benefit in doing so for our purpose, hence we stick with the Cartesian representation in $\vec d$ space.

\subsection{Coupling to an external laser field and equations of motion} \label{sec:couplingandEOMs}
If an explicitly time-dependent driver, e.g., a laser, is added to the original Hamiltonian $\hat H$ 
 the hopping elements in the Hamiltonian $H_{\alpha\beta}(\Delta\vec R)$ in \eqref{eq:tbH} become time-dependent too, and the 
 time-dependent Schrödinger equation reads
 \beq \imagi\partial_t \ket{\Psi(t)}=\sum_{\Delta\vec R\alpha\beta} H_{\alpha\beta}(\Delta\vec R,t) \sum_{\vec R} \ket{\phi_{\vec R\alpha}} \braket{\phi_{\Delta\vec R+\vec R,\beta}}{\Psi(t)}. \label{eq:tdsetb}\eeq Care has to be exercised to ensure that the tight-binding hopping Hamiltonian leads to gauge-invariant results with respect to the coupling to external fields \cite{PhysRevB.51.4940}. The fact that the usual Peierls substitution obeys this gauge-invariance of length and velocity gauge for finite SSH chains was shown explicitly in \cite{JuerssSSH2019}.

If the Bloch ansatz is chosen properly, the coupling to a laser field in dipole approximation amounts to the replacement 
\beq \vec k \quad \longrightarrow \quad \vec k(t)=\vec k + \vec A(t) \eeq
in the Bloch-Hamiltonian, where $\vec A(t)$ is the vector potential. In Appendix~\ref{appendix:properblochansatz}, we show exemplarily for the SSH chain that this does not hold for the ``wrong'' Bloch ansatz without the $\boldsymbol{\tau}_\alpha$ in \eqref{eq:Blochlikeansatz} sometimes adopted in the literature. Moreover, the ansatz without the $\boldsymbol{\tau}_\alpha$ in \eqref{eq:Blochlikeansatz} complicates the calculation of the correct velocity or current responsible for HHG because both are then not simply proportional to the expectation value of $ \boldsymbol{\nabla}_{\vec k} \mathbf{H} (\vec k)$.

Given that we choose the proper Bloch ansatz \eqref{eq:Blochlikeansatz}, the time-dependent Schrödinger equation \eqref{eq:tdsetb} boils down to
\beq \imagi \dot{ \mathbf{C}} (\vec k,t) = \mathbf{H}(\tilde{\vec d}) \mathbf{C}(\vec k,t),\qquad \tilde{\vec d}=\vec d[\vec k(t)] . \label{eq:eom1}\eeq
Typically, the propagation for a given $\vec k$ starts at $t=0$ with the electron in the valence band, $\mathbf{C}(\vec k,0)=\mathbf{C}_-(\vec k)$. The expectation value of the velocity in direction $j=1,2, \ldots D$ is given by (see Sec.\ \ref{sec::velocity})
\beq v_j(\vec k, t) = \mathbf{C}^\dagger(\vec k,t)  \partial_{k_j} \mathbf{H} (\tilde{\vec d}) \mathbf{C}(\vec k,t) . \label{eq:expforv} \eeq
HHG spectra can then be calculated by Fourier-transforming the $\vec k$-integrated acceleration $\dot{v}_j(t)$ \cite{bandrauk_quantum_2009,baggesen_dipole_2011,bauer_computational_2017}, where \beq v_j(t) = \frac{V_{\mathrm{cell}}}{(2\pi)^D}\int_{\mathrm{BZ}}\diff^D k\, v_j(\vec k, t) \label{eq:vkintegrated}. \eeq

While equation~\eqref{eq:eom1} is, from the numerical point of view, most convenient to solve and, in fact, is used to obtain reference results for HHG spectra, it is not yet suited to gain insight into the HHG process, let alone to identify nontrivial topological effects. We therefore proceed and expand in quasistatic states,
\beq \mathbf{C}(\vec k,t) = \alpha_{\vec k-}(t) \mathbf{C}_-(\tilde{\vec d}) + \alpha_{\vec k+}(t) \mathbf{C}_+(\tilde{\vec d}), \label{eq:expinqsstate} \eeq
where the adiabatic states $ \mathbf{C}_{\pm}(\tilde{\vec d})$ fulfill
\beq E_\pm(\tilde{\vec d}) \mathbf{C}_\pm (\tilde{\vec d}) = \mathbf{H}(\tilde{\vec d}) \mathbf{C}_\pm(\tilde{\vec d}). \eeq
The equation of motion for $\alpha_{\vec k\pm}(t)$ follows from \eqref{eq:eom1} and reads
\begin{widetext}
\begin{align}
\imagi \twovec{\dot{\alpha}_{\vec k+}(t)}{\dot{\alpha}_{\vec k-}(t)} &= \twobytwo{E_+(\tilde{\vec d})- \imagi\mathbf{C}^\dagger_+(\tilde{\vec d})\cdot\dot{\mathbf{C}}_+(\tilde{\vec d})}{-\imagi\mathbf{C}^\dagger_+(\tilde{\vec d})\cdot\dot{\mathbf{C}}_-(\tilde{\vec d})}{- \imagi\mathbf{C}^\dagger_-(\tilde{\vec d})\cdot\dot{\mathbf{C}}_+(\tilde{\vec d})}{E_-(\tilde{\vec d})- \imagi\mathbf{C}^\dagger_-(\tilde{\vec d})\cdot \dot{\mathbf{C}}_-(\tilde{\vec d})} \twovec{{\alpha}_{\vec k+}(t)}{{\alpha}_{\vec k-}(t)}.
\end{align}
\end{widetext}
Now we could perform a gauge transformation of the eigenvectors $\mathbf{C}_\pm(\tilde{\vec d})$ in \eqref{eq:generalpossibleCplus} and \eqref{eq:generalpossibleCminus} in order to fulfill the so-called parallel-transport gauge \cite{vanderbiltbook,topinsshortcourse} condition for the Berry connection, i.e., $\imagi\mathbf{C}^\dagger_\pm(\tilde{\vec d})\cdot\dot{\mathbf{C}}_\pm(\tilde{\vec d})=0$. However, this gauge transformation will affect the cross terms $\imagi\mathbf{C}^\dagger_\pm(\tilde{\vec d})\cdot\dot{\mathbf{C}}_\mp(\tilde{\vec d})$ such that, in the end, the same equations are found for observables such as velocities or currents. Hence we keep \eqref{eq:generalpossibleCplus}, \eqref{eq:generalpossibleCminus} and do not assume $\imagi\mathbf{C}^\dagger_\pm(\tilde{\vec d})\cdot\dot{\mathbf{C}}_\pm(\tilde{\vec d})=0$. The diagonal elements can be transformed away by the substitution
\beq {\alpha}_{\vec k\pm}(t) =  {\eta}_{\vec k\pm}(t) \,\eulere^{-\imagi \int^t(E_\pm(\tilde{\vec d}) -\imagi\mathbf{C}^\dagger_\pm(\tilde{\vec d})\cdot\dot{\mathbf{C}}_\pm(\tilde{\vec d}) )\,\diff t' }, \label{eq:fromalphatoeta}\eeq
where, under the integral in the exponent, $\tilde{\vec d}=\vec d[\vec k(t')]$, 
leading to
\begin{widetext}
\begin{align}
\imagi \twovec{\dot{\eta}_{\vec k+}(t)}{\dot{\eta}_{\vec k-}(t)} &= \twobytwo{0}{-{\cal A}_{+-}(\tilde{\vec d})\,\eulere^{\imagi \int^t(\Delta E(\tilde{\vec d}) -\Delta {\cal A}(\tilde{\vec d}) ) \,\diff t' }}{-{\cal A}_{-+}(\tilde{\vec d})\,\eulere^{-\imagi \int^t(\Delta E(\tilde{\vec d}) -\Delta {\cal A}(\tilde{\vec d}) ) \,\diff t' }}{0} \twovec{{\eta}_{\vec k+}(t)}{{\eta}_{\vec k-}(t)} .\label{eq:eomforetaingeneral}
\end{align}
\end{widetext}
Here,
\begin{align}
\Delta E(\tilde{\vec d}) &= E_+(\tilde{\vec d}) - E_-(\tilde{\vec d}) = 2|\tilde{\vec d}| = 2 \tilde d
\end{align}
is the energy difference between conduction and valence band,
\begin{align}
 {\cal A}_{\pm\pm}(\tilde{\vec d}) &=  \imagi\mathbf{C}^\dagger_\pm(\tilde{\vec d})\cdot\dot{\mathbf{C}}_\pm(\tilde{\vec d})
 \end{align}
 are the intraband Berry connections,
 \begin{align}
{\cal A}_{\pm\mp}(\tilde{\vec d}) &=  \imagi\mathbf{C}^\dagger_\pm(\tilde{\vec d})\cdot\dot{\mathbf{C}}_\mp(\tilde{\vec d})
\end{align}
are the interband Berry connections, and
\begin{align}
\Delta {\cal A}(\tilde{\vec d}) &= {\cal A}_{++}(\tilde{\vec d}) - {\cal A}_{--}(\tilde{\vec d}).
\end{align} 
The functions ${\eta}_{\vec k\pm}(t)$ are invariant under structure-gauge transformations \eqref{eq:structuregaugetrafo}.
In terms of $\tilde{\vec d}$, we find for the Berry connections, using $\abl{}{t}=\sum_{j=1}^D\dot{k}_j \partial_{k_j}$ and 
\begin{align}
{\cal D}_j(a,b) &= a\partial_{k_j}b- b\partial_{k_j}a,
\end{align}
\begin{align}
{\cal A}_{++}(\tilde{\vec d}) &  =  -\frac{\sum_j \dot k_j {\cal D}_j(\tilde{d}_x,\tilde{d}_y)}{2\tilde d(\tilde d+\tilde{d}_z)}, \\
{\cal A}_{--}(\tilde{\vec d}) &=   -\frac{\sum_j \dot k_j{\cal D}_j(\tilde{d}_x,\tilde{d}_y)}{2\tilde d(\tilde d-\tilde{d}_z)} = {\cal A}_{++}(-\tilde{\vec d}), \\
{\cal A}_{+-}(\tilde{\vec d}) & = \frac{\sum_j \dot k_j \left( {\cal D}_j(\tilde{d}_x,\tilde{d}_y)  + \imagi {\cal D}_j(\tilde{d}_z,\tilde d) \right) }{2\tilde d \tilde d_\perp}, \\
{\cal A}_{-+}(\tilde{\vec d}) &=  {\cal A}^*_{+-}(\tilde{\vec d}), \\
\Delta {\cal A}(\tilde{\vec d}) &= \frac{\tilde{d}_z}{\tilde d \tilde d^2_\perp}\sum_j \dot k_j{\cal D}_j(\tilde{d}_x,\tilde{d}_y).
\end{align}

\subsection{Electron velocity} \label{sec::velocity}
 With the proper Bloch ansatz, the velocity operator for an initial $\vec k$ and in direction $j$ becomes the $2\times 2$-matrix 
\begin{align} \mathbf{v}_j(\tilde{\vec d}) &= \partial_{k_j} \mathbf{H}(\tilde{\vec d}) =  \partial_{k_j} \tilde{\vec d} \cdot\boldsymbol{\sigma}.
\end{align}
Since $\tilde{\vec d}=\vec d[\vec k(t)]=\vec d[\vec k + \vec A(t)]$ we understand that $\partial_{k_j} \mathbf{H}(\tilde{\vec d})=\partial_{k_j} \mathbf{H}(\vec k)\bigr|_{\vec k + \vec A(t)}$.
The expectation value for the velocity in direction $j$ of a laser-driven electron starting at lattice momentum  $\vec k$ thus is
\begin{align}
v_j(\vec k,t) &=  \mathbf{C}^\dagger (\vec k,t) \mathbf{v}_j(\tilde{\vec d})  \mathbf{C}(\vec k,t) \label{eq:veloinjdir} \\
&= v_j^{--}(\vec k,t) + v_j^{++}(\vec k,t) + v_j^{-+}(\vec k,t) + v_j^{+-}(\vec k,t)
\end{align}
where, using \eqref{eq:expinqsstate} and \eqref{eq:fromalphatoeta},
\begin{align}
 v_j^{--}(\vec k,t) &=  |{\eta}_{\vec k-}(t)|^2 \mathbf{C}^\dagger_-(\tilde{\vec d}) \mathbf{v}_j(\tilde{\vec d}) \mathbf{C}_-(\tilde{\vec d}), \\
v_j^{++}(\vec k,t) &=  |{\eta}_{\vec k+}(t)|^2  \mathbf{C}^\dagger_+(\tilde{\vec d}) \mathbf{v}_j(\tilde{\vec d}) \mathbf{C}_+(\tilde{\vec d}),
\end{align}
and
\begin{align}
v_j^{-+}(\vec k,t) &= {\eta}^*_{\vec k-}(t) {\eta}_{\vec k+}(t) \,\eulere^{-\imagi \int^t(\Delta E(\tilde{\vec d}) -\Delta {\cal A}(\tilde{\vec d})) \,\diff t' } \nonumber \\
& \qquad \times  \mathbf{C}^\dagger_-(\tilde{\vec d}) \mathbf{v}_j(\tilde{\vec d})\mathbf{C}_+(\tilde{\vec d}),\\
v_j^{+-}(\vec k,t) &= \left[v_j^{-+}(\vec k,t)\right]^*.
\end{align}
After a cumbersome but straightforward calculation, we obtain
\begin{align}
 v_j^{\pm\pm}(\vec k,t) &=\pm |{\eta}_{\pm\vec k}(t)|^2 \partial_{k_j} \tilde d, \label{eq:intrabandvelos} \\
v_j^{-+}(\vec k,t) &= -{\eta}^*_{\vec k-}(t) {\eta}_{\vec k+}(t) \,\eulere^{-\imagi \int^t\left[2\tilde d - \Delta {\cal A}(\tilde{\vec d}) \right] \,\diff t' }\nonumber \\
& \qquad  \times \frac{1}{\tilde d_\perp}\left[ {\cal D}_j(\tilde d_z,\tilde d) + \imagi {\cal D}_j(\tilde d_x,\tilde d_y)  \right],  \label{eq:interbandveloplusminusk}\\
v_j^{+-}(\vec k,t) &=  \bigl[v_j^{-+}(\vec k,t)\bigr]^*. \label{eq:vplusminusgeneral}
\end{align}
The velocity components \eqref{eq:intrabandvelos} do not mix ${\eta}_{+\vec k}(t)$ and ${\eta}_{-\vec k}(t)$ and hence might be called intraband velocities.
We see that these two intraband velocities are in opposite directions because of the symmetry in the dispersion relation $E_\pm(\vec k)=\pm d(\vec k)$. The weighting factors $|{\eta}_{\pm\vec k}(t)|^2$ account for the populations of the two bands.
Berry curvature effects come into play through the velocity contributions \eqref{eq:interbandveloplusminusk}, \eqref{eq:vplusminusgeneral} that do mix ${\eta}_{+\vec k}(t)$ and ${\eta}_{-\vec k}(t)$, and thus might be called interband velocities. A similar observation has been made in \cite{chacon_observing_2018,YueGaarde2020} for the relation between dipole transition matrix elements and the Berry curvature. The total $\vec k$-resolved velocity expectation value in direction $j$ is
\beq   v_j(\vec k,t) = \left( |{\eta}_{\vec k+}(t)|^2 -  |{\eta}_{\vec k-}(t)|^2 \right) \partial_{k_j} \tilde d + 2 \Re v_j^{-+}(\vec k,t) . \label{eq:vjkt} \eeq
The first term is the expected group velocity, the second term is the anomalous velocity, including all topological effects.
Integration over the Brillouin zone yields the total velocity in direction $j$
\beq  v_j(t) =  \frac{V_{\mathrm{cell}}}{(2\pi)^D} \int_{\mathrm{BZ}} \diff^D k \,  v_j(\vec k,t),\eeq
which is proportional to the current if the correct Bloch ansatz is chosen.

\subsection{Lewenstein-like model for high-harmonic generation in two-band systems} \label{sec:lewensteinmodelling}
 The intraband velocities have the simple form of the populations $|{\eta}_{\vec k\pm}(t)|^2$ in the respective bands times the corresponding group velocities $\left[\partial_{k_j} E_\pm(\vec k)\right]_{\vec k + \vec A(t)}=\pm \partial_{k_j} \tilde d$.  

Harmonic generation by the interband velocity $2 \Re v_j^{-+}(\vec k,t)$ might be viewed similar to the three-step HHG in atomic gas targets \cite{vampa_merge_2017}: (i) the electron makes a vertical transition from the valence to the conduction band, (ii) the electron oscillates in the conduction band, and (iii) the electron recombines into the valence band upon emission of a photon whose energy equals the band gap at the $\vec k$-point where recombination takes place. In that picture, it is assumed that the laser field does not strongly affect the band structure so that harmonics up to the maximum field-free band gap are expected.

Intuitively, one might think that harmonic spectra calculated from the intraband velocity alone do not show a plateau up to the maximum energy gap but only low-order harmonics. However, this is not true, as the energy gap $2d(\vec k)$ enters the expression for $\eta_{\vec k+}(t)$, see equation \eqref{eq:etakpluslewen}.

The initial conditions, describing a fully occupied valence band and an empty conduction band, read $ \alpha_{\vec k+}(0)=\eta_{\vec k+}(0)=0$ and $\alpha_{\vec k-}(0) =\eta_{\vec k-}(0)=1$. The assumption in the Lewenstein paper on gas HHG \cite{LewensteinPhysRevA.49.2117} that depletion of the population in the electronic ground state is negligible in the parameter regime of interest translates to $\eta_{\vec k-}(t)\simeq 1$. Note that the assumption $\alpha_{\vec k-}(t)\simeq 1$ is not valid because of the complex phase that $\alpha_{\vec k-}(t)$ accumulates (even without laser). 
With $\eta_{\vec k-}(t)\simeq 1$ at all times we find\ \footnote{Here, it is again understood that the correct time-dependencies have to be employed, i.e., ${\cal A}_{+-}(\tilde{\vec d}) = {\cal A}_{+-}(\vec d [\vec k(t')])$ and, in the exponent, $2\tilde d -\Delta {\cal A}(\tilde{\vec d}) =2d[\vec k(t'')] -\Delta {\cal A}(\vec d[\vec k(t'')]) $.} 
\begin{align} \eta_{\vec k+}(t) &= \imagi\int^t {\cal A}_{+-}(\tilde{\vec d})\,\eulere^{\imagi \int^{t'}[2\tilde d -\Delta {\cal A}(\tilde{\vec d}) ] \,\diff t'' } \,\diff t' .  \label{eq:etakpluslewen}
\end{align}
As a consequence, eq.\ \eqref{eq:vjkt} becomes
\beq   v_j(\vec k,t) \simeq \left( |{\eta}_{\vec k+}(t)|^2 - 1 \right) \partial_{k_j} \tilde d + 2 \Re v_j^{-+}(\vec k,t)  \label{eq:vjktlewen} \eeq
with
\begin{align}
v^{-+}_j(\vec k,t) &\simeq      \tilde{\cal A}_{-+}^j(\tilde {\vec d}) \int^t   {\cal A}_{+-}(\tilde {\vec d})   \eulere^{-\imagi S(\vec k, t', t)} \,\diff t' \label{eq:vplusminus0}
\end{align}
where
\beq \tilde{\cal A}_{-+}^j(\tilde{\vec d}) = \tilde{\cal A}_{+-}^{j*}(\tilde{\vec d}) =\frac{1}{\tilde d_\perp} \left[ {\cal D}_j(\tilde d_x,\tilde d_y)   -\imagi {\cal D}_j(\tilde d_z,\tilde d)  \right] \eeq
and the action $S(\vec k, t', t)$ is
\beq S(\vec k, t', t) = \int_{t'}^t\left[2 \tilde d -\Delta {\cal A}(\tilde{\vec d}) \right] \,\diff t'' . \label{eq:ouraction} \eeq
The set of equations \eqref{eq:etakpluslewen}--\eqref{eq:ouraction} for the ($\vec k$-resolved) electron velocity is the main result of this work. It provides an explicit expression of the total velocity in terms of $\vec d(\vec k)$ defining the system under consideration. Topological effects are included via the interband velocity \eqref{eq:vplusminus0}. Hence, topologically interesting changes in the chirality of the current (e.g., clockwise or counter-clockwise around a certain $\vec k$-point) can be analyzed. One may also ``reverse engineer'' a topologically interesting system by defining $\vec d$ such that the interband current yields the desired (laser-driven) electron dynamics. Of course, a system designed in such a way may correspond to weird hoppings in position space (an example being the Qi-Wu-Zhang toy model \cite{QiWuZhang06,topinsshortcourse} with a simple $\vec d$ but complicated position-space hoppings). 

Note that we did neither apply a single-band approximation or semi-classical wave-packet dynamics nor is our result restricted to particular dimensions. As a consequence, the velocity \eqref{eq:vjktlewen} is more general than the commonly employed $\vec v = \boldsymbol{\nabla}_{\vec k} E(\vec k) - \dot{\vec k}\times \boldsymbol{\Omega(\vec k)}$ \cite{PhysRevB.59.14915,Gosselin_2006}, where $\boldsymbol{\Omega(\vec k)}$ is the Berry curvature.

The $\vec k$-integrated interband velocity that will, after Fourier transformation, contribute to interband HHG, reads $2\Re v^{-+}_j(t)$ with
\begin{align}
v^{-+}_j(t) &\simeq  \frac{V_{\mathrm{cell}}}{(2\pi)^D} \int_{\mathrm{BZ}} \diff^D k \,  \tilde{\cal A}_{-+}^j(\tilde {\vec d})  \nonumber \\
& \quad \times \int^t   {\cal A}_{+-}(\tilde {\vec d})   \eulere^{-\imagi S(\vec k, t', t)} \,\diff t'. \label{eq:vplusminuskintegrated}
\end{align}
The structure of this expression is the same as for the dipole in the celebrated Lewenstein paper on HHG in gases \cite{LewensteinPhysRevA.49.2117} so that one could embark on transferring all steps outlined there to solids. 

The analogy between the three-step model in gas HHG and interband HHG in solids is well known \cite{vampa_merge_2017}. In \cite{PhysRevX.7.021017}, a mixed Wannier-Bloch representation is employed for the valence band (Wannier) and conduction band (Bloch), which elucidates the similarity between gas-phase and solid HHG most clearly because Wannier functions are localized in position space (like the ground state wave function in atomic HHG). In that way one can follow where electrons start and recombine in position space. 
However, the for all practical purposes crucial differences between gas-phase HHG and HHG in solids are the following. First, the action in the gas-phase HHG is simple and reads $S(\vec p, t', t) = \int_{t'}^t\{ [\vec p + \vec A(t'') ]^2/2 + I_p \} \,\diff t''$ where $\vec p$ is the canonical momentum of the electron and $I_p$ is the ionization potential of the atom. Instead, the functional dependence of the action \eqref{eq:ouraction} on $\vec k$ is rather involved even for the simplest model solids so that the time-integral---after insertion of $\vec k(t)$---cannot be performed analytically. 
Second, the dipole transition matrix elements in atomic HHG are rather simple whereas the interband couplings in \eqref{eq:eomforetaingeneral} and the interband velocity \eqref{eq:interbandveloplusminusk} expressed explicitly in terms of $\vec d$ are rather involved. By making approximations to these couplings one may easily sweep topological effects under the carpet, as was also pointed out recently in \cite{YueGaarde2020}. The main objective of our paper is to provide explicit, analytical expressions for the velocity, including all topological effects and without any approximations besides tight-binding and the restriction to two bands.
Further, we note in passing that the $\vec k$-integration in the solid-state result \eqref{eq:vplusminuskintegrated} is performed because all $\vec k$ states in the valence band are initially populated. Hence, HHG in solids, described by eq.\ \eqref{eq:vplusminuskintegrated}, includes many-electron effects such as the interference of the radiation emitted by ``individual'' electrons while interaction between the electrons is not taken into account. Instead, the $\vec p$ integration in gas-phase HHG arises already for a single active electron.

In the Lewenstein paper on gas HHG \cite{LewensteinPhysRevA.49.2117}, the integration over the electron's canonical momentum $\vec p$ is performed using saddle-point integration. The beauty is that the saddle-point integration there is not just a mathematical trick but allows for an intuitive interpretation: only those semi-classical trajectories contribute to HHG that start at the ionization time $t'$ at the origin (where the parent ion is located) and return to the origin at the recombination time $t$. This makes sense because recombination can only take place at the position of the ion. 
We may try to proceed analogously to the Lewenstein paper and perform in \eqref{eq:vplusminuskintegrated} the integration with respect to $\vec k$ by searching for stationary $\vec k_\mathrm{st}(t,t')$ that fulfill
\beq \boldsymbol{\nabla}_{\vec k}  S(\vec k, t', t) = \boldsymbol{0} .\eeq
If there was not the $\Delta {\cal A}(\vec d)$-term in the action \eqref{eq:ouraction} we would obtain
\[ \boldsymbol{0} =\boldsymbol{\nabla}_{\vec k}  S(\vec k, t', t) = 2\int_{t'}^t \boldsymbol{\nabla}_{\vec k} \tilde d   \,\diff t'' =  -2\int_{t'}^t {\vec v}^{--}(\vec k,t'') \,\diff t'' \]
from which follows
\[ \vec r(\vec k,t) - \vec r(\vec k,t') = \boldsymbol{0} \]
where we have defined formally a position \[ \vec r(\vec k,t) = \int^t {\vec v}^{--}(\vec k,t'') \,\diff t''.\]
Hence, we find formally the same result as for HHG in atoms: the semi-classical electron trajectory returns to its starting point {\em in position space}. This semi-classical viewpoint also emerged in the studies of HHG in solids based on optical Bloch equations \cite{vampa_merge_2017}. However, incomplete returns also contribute to HHG in solids \cite{PhysRevLett.124.153204}. 

In atomic HHG, the saddle-point equation $ \boldsymbol{\nabla}_{\vec p} S(\vec p, t', t) = \boldsymbol{0}$ can be easily evaluated and solved explicitly for the stationary momentum $\vec p_\mathrm{st}(t,t')$. Due to the more involved dispersion relations and the possible presence of the Berry term $\Delta {\cal A}(\vec d)$ in the case of solids, it is not possible to find explicit expressions for $\vec k_\mathrm{st}(t,t')$. If the laser field is sufficiently weak such that $\vec A(t)$ is much smaller than the dimensions of the Brillouin zone, one may expand the integrand in $S(\vec k, t', t)$
 up to $\vec A^2(t'')$ and perform the time integral over $t''$. In that way it is possible to factorize $\vec k$-dependence and time-dependence in the action. Yet, the result will still have a too complicated dependence on $\vec k$ to find explicit expressions for $\vec k_\mathrm{st}(t,t')$. However, a graphical or numerical solution would yield, for given excitation and recombination times $t'$, $t$, the dominating $\vec k$, which might be useful for the analysis or interpretation of numerically obtained results. The numerical calculation of entire HHG spectra in this way is not recommended, as it would be much less efficient than simply solving the differential equation \eqref{eq:eom1} numerically.

\section{Results} \label{sec:results}
We now test the validity of our theory by applying it to two prime examples of model systems described by $2\times 2$ Bloch-Hamiltonians: the SSH chain and the Haldane model.

\subsection{SSH case} \label{sec:resultsSSH}
In Appendix \ref{appendix:properblochansatz}, we introduce the position-space representation of the SSH Hamiltonian and derive the Bloch-Hamiltonian
\begin{align} \mathbf{H}(k) &= \twobytwo{0}{v\eulere^{\imagi k/2} + w \eulere^{-\imagi k/2}}{v\eulere^{-\imagi k/2} + w \eulere^{\imagi k/2}}{0} \nonumber \\
&= (v+w)\cos(k/2) \boldsymbol{\sigma}_x + (w-v)\sin(k/2) \boldsymbol{\sigma}_y
\end{align}
such that the velocity operator is indeed $\partial_k H(\tilde{\vec d})$, and the current is proportional to it. We choose real $v$ and $w$, and a lattice constant $a=1$. Obviously,
\beq \vec d = \threevec{(w+v)\cos(k/2)}{(w-v)\sin(k/2)}{0}, \eeq
the dispersion relation is
\beq E_\pm(k)= \pm d=\pm\sqrt{w^2+v^2+2wv \cos k}, \eeq
and $ d = d_\perp$. 
There is only one direction $j=1$, and the driver $\vec A$ is necessarily parallel to it. 
Equation \eqref{eq:etakpluslewen} becomes in this case
\begin{align*} \eta_{k+}(t) &= \frac{\imagi (w^2-v^2)}{4}\int^t \diff t'\, \frac{\dot A(t')}{E^2_+(k+A(t'))}\\
& \qquad \qquad\times\eulere^{\imagi \int^{t'}2 E_+(k+A(t''))   \,\diff t'' } , 
\end{align*}
and eq.\ \eqref{eq:vplusminus0} reads
\begin{align}
v^{-+}(k,t) &\simeq   \frac{(w^2-v^2)^2}{8E_+(k+A(t))}\int^t  \diff t' \, \frac{\dot A(t')}{E_+^2(k+A(t'))}   \nonumber\\
 & \qquad \qquad\times \eulere^{-\imagi \int^t_{t'}2E_+(k+A(t'')) \,\diff t'' } 
. \label{eq:SSHreloadedvminplus3}
\end{align}
This interband velocity is inserted into \eqref{eq:vjktlewen}. 

It is known that the topological phase transition of the SSH chain occurs at $w=v$, with $w>v$ giving rise to the nontrivial topological phase, with edge states in finite SSH chains \cite{topinsshortcourse}. However, both the intraband velocity $|{\eta}_{\vec k+}(t)|^2\partial_{k} \tilde d$ and the interband velocity \eqref{eq:SSHreloadedvminplus3} are proportional to $(w^2-v^2)^2$, i.e., completely symmetric under an exchange $w \leftrightarrow v$ so that there is no way to distinguish the trivial and the nontrivial topological phase via HHG in SSH bulk. In fact, for periodic boundary conditions the dangling sites for $w>v$ in a finite chain pair-up, and the velocity expectation value should be invariant under the exchange $v \leftrightarrow w$. In contrast, in finite systems, where the edge states show up explicitly in the topologically nontrivial SSH phase $w>v$, huge differences in the HHG yield between trivial and nontrivial topological phase are observed \cite{bauer_high-harmonic_2018,DrueekeRobustness2019,JuerssSSH2019}. 

The $k$-integration required in \eqref{eq:vplusminuskintegrated} is performed numerically by sampling the Brillouin zone $[-\pi,\pi[$ with $N_k$ equidistant $k$ values. The result should be the same as that for a calculation in position space with $N=N_k$ unit cells and periodic boundary condition.

\begin{figure}
\includegraphics{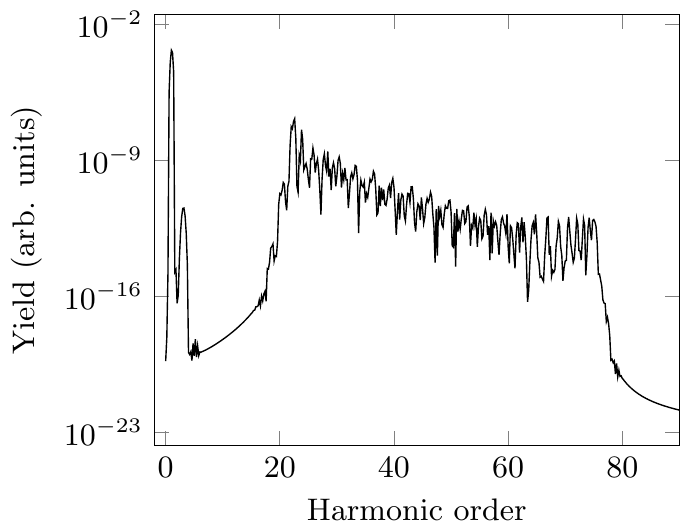}
\caption{HHG spectrum for a SSH chain in a laser field with vector potential \eqref{eq:laserfield} (SSH and laser parameters are given in the text). 
\label{fig1}}
\end{figure}

Fig. \ref{fig1} shows the HHG spectrum for the SSH chain with $v=-\eulere^{-1.7}\simeq -0.1827$, $w=-\eulere^{-2.3}\simeq -0.1003$ in a laser field of the form
\beq A(t)= A_0 \sin^2\left( \frac{\omega t}{2 n_\mathrm{cyc}} \right) \sin\omega t  \label{eq:laserfield}\eeq
with $A_0=0.1$, $\omega=0.0075$, $n_\mathrm{cyc}=5$,
calculated by Fourier-transforming the first time-derivative of the velocity expectation value (i.e., the acceleration). We have checked that the calculations (i) directly in position space for a chain with $N= 50$ unit cells and periodic boundary conditions (see Appendix \ref{appendix:properblochansatz}), (ii) according eqs.\ \eqref{eq:eom1}, \eqref{eq:expforv}, and \eqref{eq:vkintegrated} (with the $k$-integral replaced by a discrete sum over $N_k=50$ equidistant $k$-values in the Brillouin zone), and (iii) according \eqref{eq:vjktlewen} all give the same spectrum, which shows that, first, the equations of motions are correct, second, that the Bloch ansatz chosen in Appendix \ref{appendix:properblochansatz} is consistent with the velocity operator $\partial_k \mathbf{H}(\tilde{\vec d})$, and third, that the assumption of negligible depletion, i.e., $\eta_{k-}(t)\simeq 1$ is valid. The HHG spectrum displays the known features \cite{JuerssSSH2019} of rapidly dropping low-order  harmonics, followed by a plateau of emission in the photon energy interval $[\min(2d),\max(2d)]$. 

\bigskip

\subsection{Haldane case} \label{sec:resultsHaldane}
The $2\times 2$ Bloch-Hamiltonian for the Haldane model is derived in Appendix \ref{appendix:haldane} and reads 
\begin{widetext}
\begin{align} \label{eq:haldane-bloch-hamiltonian}
	\mathbf{H}(\boldsymbol{k}) &= \twobytwo%
	{M + t_2 \sum_n \eulere^{-\imagi \boldsymbol{k} \cdot \boldsymbol{g}_n} + t_2^* \sum_n \eulere^{\imagi \boldsymbol{k} \cdot \boldsymbol{g}_n}}%
	{t_1 \sum_n \eulere^{\imagi \boldsymbol{k} \cdot \boldsymbol{\delta}_n}}%
	{t_1 \sum_n \eulere^{-\imagi \boldsymbol{k} \cdot \boldsymbol{\delta}_n}}%
	{-M + t_2 \sum_n \eulere^{\imagi \boldsymbol{k} \cdot \boldsymbol{g}_n} + t_2^* \sum_n \eulere^{-\imagi \boldsymbol{k} \cdot \boldsymbol{g}_n}}
	\nonumber \\
	&=
	2 \Real(t_2) \sum_n \cos(\boldsymbol{k} \cdot \boldsymbol{g}_n) \boldsymbol{1} +
	t_1 \sum_n \cos(\boldsymbol{k} \cdot \boldsymbol{\delta}_n) \boldsymbol{\sigma}_x -
	t_1 \sum_n \sin(\boldsymbol{k} \cdot \boldsymbol{\delta}_n) \boldsymbol{\sigma}_y +
	\Bigl( M + 2 \Imag(t_2) \sum_n \sin(\boldsymbol{k} \cdot \boldsymbol{g}_n) \Bigr) \boldsymbol{\sigma}_z
	.
\end{align}
\end{widetext}
The real part of the next-nearest neighbor hopping amplitude \( t_2 \) shifts the energy but does not change the energy difference between both bands. 
As a consequence, the derivative of the band structure might be influenced, which changes the intraband velocity. However, in this work we choose a purely imaginary \( t_2 \). Further studies might investigate the influence of a non-vanishing real part of $t_2$.
We obtain for the $\boldsymbol{d}$-vector in \eqref{eq:2x2BlochHamil}
\begin{equation}
	\boldsymbol{d}(\boldsymbol{k}) = \threevec%
	{t_1 \sum_n \cos(\boldsymbol{k} \cdot \boldsymbol{\delta}_n)}%
	{-t_1 \sum_n \sin(\boldsymbol{k} \cdot \boldsymbol{\delta}_n)}%
	{M + 2 \Imag(t_2) \sum_n \sin(\boldsymbol{k} \cdot \boldsymbol{g}_n)},
\end{equation}
and
\begin{align}
	d_{\bot} &= \sqrt{|t_1|^2 \Bigl( 3 + 2 \sum_n \cos(\boldsymbol{k} \cdot \boldsymbol{g}_n) \Bigr)}, \\
	d &= \sqrt{d_{\bot}^2 + \Bigl( M + 2 \Imag(t_2) \sum_n \sin(\boldsymbol{k} \cdot \boldsymbol{g}_n) \Bigr)^2}
	.
\end{align}

In the following, we show exemplarily results for HHG due to laser-driven electron dynamics around the K point and the K' point for the topologically trivial and nontrivial phase.
Both for the testing of our theory and for a better understanding it is instructive to look at the contributions from specific $\vec k$-points separately.
Afterwards, an integration over the Brillouin zone is performed to obtain measurable HHG spectra.

\begin{figure}
\includegraphics{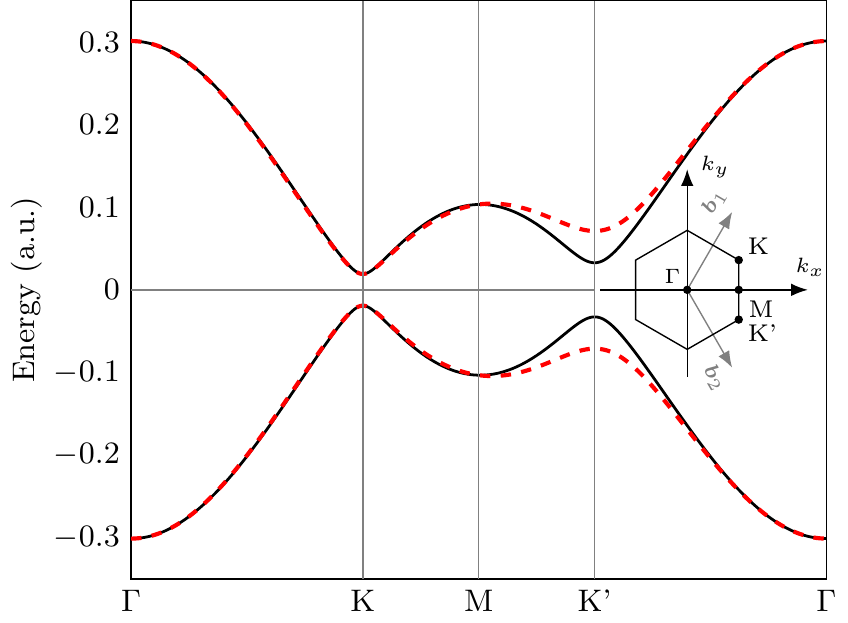}
\caption{Band structure of the Haldane model for $a=2.683$, $M=0.026$, $t_1=-0.1$, and $t_2=-0.0013\imagi$ (trivial, black) and $t_2=-0.0087\imagi$ (nontrivial, red dashed) 
\label{fig:haldane-band-and-kspace-points}}
\end{figure}

The Haldane model parameters are $a=2.683$, $M=0.026$, $t_1=-0.1$, and $t_2=-0.0013\imagi$ (trivial) and $t_2=-0.0087\imagi$ (nontrivial). The laser pulse is the same as in the SSH example \eqref{eq:laserfield} and polarized in $\Gamma$M-direction. Fig. \ref{fig:haldane-band-and-kspace-points} shows the band structure for the two parameter sets. The values for $t_2$ were chosen such that the smallest band gap, which is at the K point, is the same below and above the topological phase transition, corresponding to $\simeq 5$ times the laser frequency.

\begin{figure}
\includegraphics{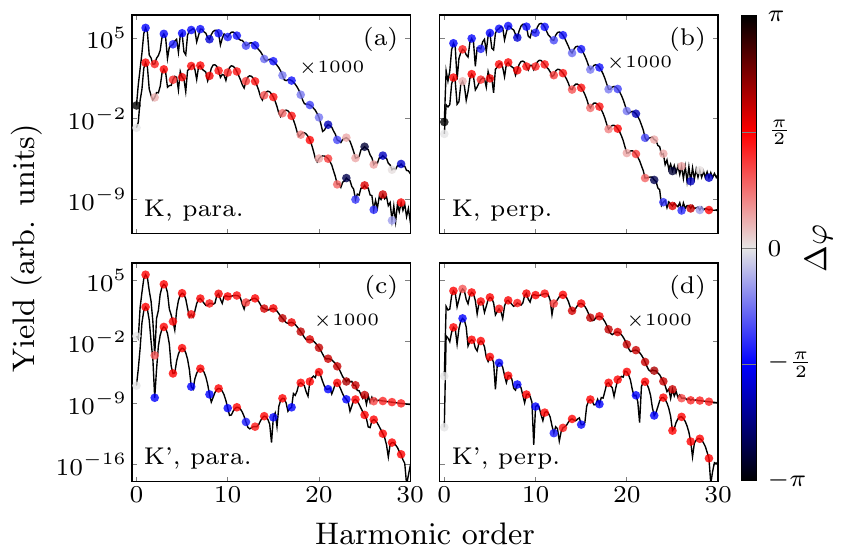}
\caption{HHG spectra generated by the electron initially at the K point (a,b) and the K' point (c,d). (a,c) HHG spectra calculated from the acceleration parallel to the polarization direction of the incoming laser field for the trivial phase (upper curve, multiplied by $1000$) and the nontrivial topological phase (lower curve). (b,d) Respective spectra from the acceleration perpendicular to the polarization direction of the incoming laser field. The phase difference \eqref{eq:helicityphase} of integer harmonics is color-coded in each panel. The Haldane model parameters are the same as in Fig.\ \ref{fig:haldane-band-and-kspace-points}. 
\label{fig:haldane-K-spectra}}
\end{figure}

The HHG spectra calculated from the acceleration of the electron initially at the K point are presented in Fig.\ \ref{fig:haldane-K-spectra}(a,b). In Fig.\  \ref{fig:haldane-K-spectra}(a) the HHG spectra in the trivial and nontrivial topological phase (Haldane model parameters as in Fig.\ \ref{fig:haldane-band-and-kspace-points}) calculated from the acceleration parallel to the incoming laser field ($\dot{v}_\parallel$) are shown. We have checked that the calculation according (i) \eqref{eq:eom1}, \eqref{eq:expforv}, according (ii) \eqref{eq:vjkt}, and (iii) assuming no depletion \eqref{eq:vjktlewen} give the same spectra. Fig. \ref{fig:haldane-K-spectra}(b) shows the corresponding spectra calculated from the acceleration perpendicular to the incoming laser field ($\dot{v}_\perp$).
The respective phase difference
\beq \Delta\varphi= \Arg(\mathrm{FFT}[\dot{v}_\parallel]) - \Arg(\mathrm{FFT}[\dot{v}_\perp])  \label{eq:helicityphase}\eeq
is color-coded in all panels of Fig.\    \ref{fig:haldane-K-spectra}. It determines the helicity of the emitted light.
The phase differences $0$ and $\pi$ (or, equivalently, $-\pi$) mean that the emitted harmonics are linearly polarized.
Other phase differences define (together with the magnitudes of the emission polarized along $x$ and $y$) the ellipticity (or helicity) of the emitted harmonics.
It is clearly seen that the phase difference of most of the harmonics flips from $-\pi/2$ in the trivial topological phase to $+\pi/2$ in the nontrivial phase, i.e. the helicity changes. Even the fundamental flips in that way, which seems in contradiction 
with the findings in \cite{Silva2019}. However, note that the polarization axes of the lasers are different in both papers. Even harmonics polarized perpendicular to the incoming laser field appear in Fig.\ \ref{fig:haldane-K-spectra}(b), with the 2nd behaving anomalously in having a helicity opposite to those of the other harmonics in the trivial phase.

\begin{figure}
\includegraphics{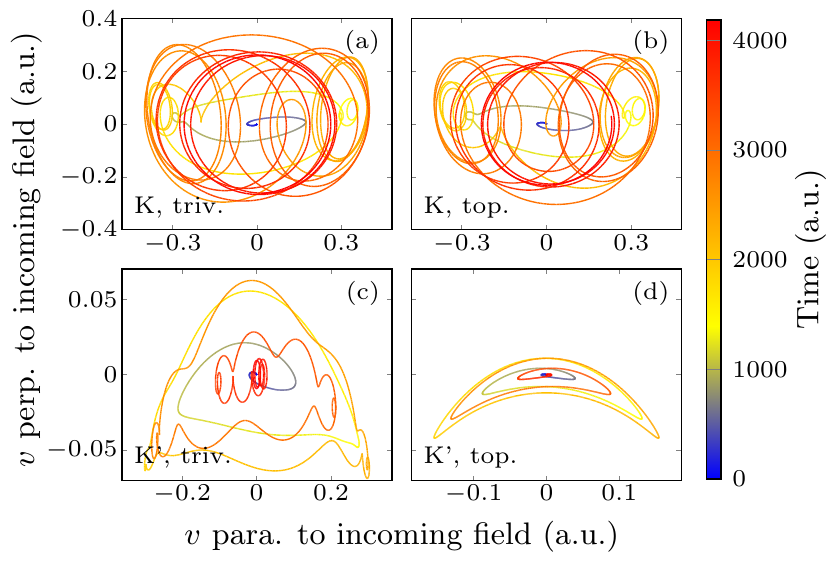}
\caption{Electron velocity for the electron initially at the K point (a,b) and K' point (c,d) in the $v_x,v_y$ (i.e., $v_\parallel,v_\perp$) plane in the topologically trivial phase (a,c) and the nontrivial phase (b,d). Haldane model parameters as in Fig.\ \ref{fig:haldane-band-and-kspace-points}. Time is color-coded. 
\label{fig:trajectories-at-K-point}}
\end{figure}

Fig. \ref{fig:trajectories-at-K-point}(a,b) shows the actual electron trajectories for the electron starting from the K point in the $v_x,v_y$ plane (i.e., $v_\parallel,v_\perp$ plane) in the trivial and nontrivial phase, respectively. The time is color-coded. It is clearly seen that the orientation of the trajectory changes from clockwise in the trivial phase to counter-clockwise in the nontrivial topological phase. Note that the velocity components $v_\parallel,v_\perp$ at the K point are similar in magnitude despite the linear polarization of the incoming pulse along $v_\parallel$. This leads to a particularly high ellipticity of the emitted harmonics and even a helicity flip of the fundamental when passing the phase transition.

\begin{figure}
\includegraphics{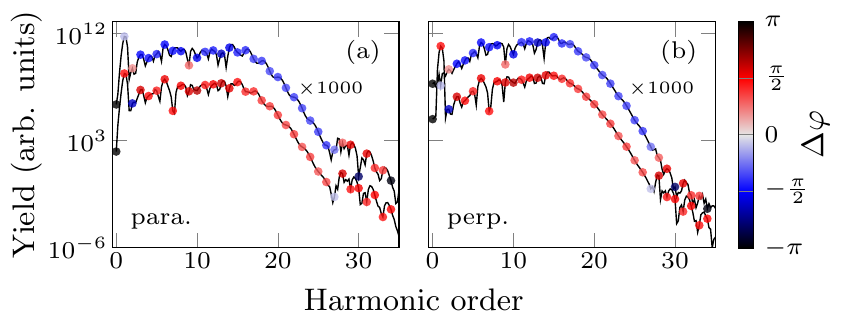}
\caption{ HHG spectra calculated from the total electron velocity, integrated over the first Brillouin zone. (a) HHG spectra calculated from the acceleration parallel to the polarization direction of the incoming laser field for the trivial phase (upper curve, multiplied by $1000$) and the nontrivial topological phase (lower curve). (b) Respective spectra from the acceleration perpendicular to the polarization direction of the incoming laser field. The phase difference \eqref{eq:helicityphase} of integer harmonics is color-coded in each panel. Haldane model parameters as in Fig.\ \ref{fig:haldane-band-and-kspace-points}. 
\label{fig:haldane-bz-spectra}}
\end{figure}

The corresponding results for the K' point are shown in Figs.\ \ref{fig:haldane-K-spectra}(c,d) and \ref{fig:trajectories-at-K-point}(c,d). The HHG spectra in Fig.\ \ref{fig:haldane-K-spectra}(c,d) differ more in shape than those for the K point because the band gaps at the K' point in the trivial and the nontrivial topological phase differ significantly (see Fig.\ \ref{fig:haldane-band-and-kspace-points}). This is why, in the nontrivial topological phase (where the band gap is larger), the characteristic band-gap dip around harmonic order 13 appears. To the left of the dip, the harmonics roll off exponentially, to the right of the dip the harmonics plateau starts to form (at higher laser intensity it would broaden). The helicity flip at the K' point is also very different from the K point. The phase difference of all harmonics is $+\pi/2$ in the trivial phase, and almost every other harmonic flips in the nontrivial topological phase (the fundamental does not flip, the 2nd harmonic does, 3rd, 4th, and 5th do not flip, the 6th does, etc.).

The velocities of the laser-driven electron that starts from the K' point for the trivial and the nontrivial topological phase are shown in Fig.\ \ref{fig:trajectories-at-K-point}(c) and (d), respectively. Note that the velocity components in perpendicular direction are much smaller than at the K point. The electron dynamics is very much aligned along the laser polarization direction. There is no switch from clockwise to counter-clockwise electron motion below and above the topological phase transition at the K' point. The motion is counter-clockwise in both cases. The trajectory looks more regular in the nontrivial topological phase, with the electron returning to zero velocity after the laser pulse. This is because of the larger band gap at the K' point in the nontrivial topological phase for the choice of our Haldane model parameters. For an increased laser intensity the electron dynamics there would also look more ``chaotic''.

Fig. \ref{fig:haldane-bz-spectra}(a,b) shows HHG spectra calculated from the $\bm{k}$-integrated electron velocity, eq. (\ref{eq:vkintegrated}). For the numerical integration, $1500 \times 1500$ $\boldsymbol{k}$-points within the first Brillouin zone were used, which was sufficient to obtain converged results.   HHG spectra for the trivial and nontrivial topological phase for polarization directions parallel and perpendicular to the incoming laser field are shown in Fig. \ref{fig:haldane-bz-spectra}(a,b), respectively.
The phase difference \eqref{eq:helicityphase} is again color-coded.
We find that the helicity flip observed at the K point survives in the $\bm{k}$-integrated result.

\section{Summary and conclusions} \label{sec:summ}
We derived the equation for the velocity of a laser-driven electron in a two-band solid explicitly in terms of the system-specific three-vector $\vec d(\vec k)$  and the laser field's vector potential $\vec A(t)$ in dipole approximation. Besides tight-binding, dipole approximation, and negligible depletion, we did not make further assumptions such as single-band approximation, semi-classical dynamics, or simplified transition matrix elements (that may break gauge invariance or suppress topological effects). We calculated harmonic spectra by Fourier-transforming the acceleration exemplarily for the Su-Schrieffer-Heeger chain and the Haldane model in intense laser fields. While for the Su-Schrieffer-Heeger chain there was no difference in the harmonic spectra above and below the topological phase transition if periodic boundary conditions are used, the helicity of the harmonics may change in the Haldane model driven by a linearly polarized laser field. The helicity changed differently for different harmonics, depending on the initial $\vec k$ point of the electron. In the overall spectrum a helicity flip for each harmonic is observed. The complex electron dynamics was illustrated by electron trajectories in the velocity plane whose orientation (i.e., chirality) swapped from clockwise to counter-clockwise at the K point but did not swap at the K' point. Our analytical formula for the electron velocity allows to analyze and predict the laser-driven electron dynamics, for instance, whether chirality swaps are expected or not. We carefully checked that our analytical equation for the velocity leads to the same results as those obtained by solving directly the differential equations of motion either in position space or $\vec k$ space. Although we applied our theory to harmonic generation, other strong-field or few-cycle pulse effects could be studied as well, for instance laser-driven valleytronics or transient absorption spectroscopy of topologically nontrivial matter.

\section*{Acknowledgment}
C.J. acknowledges financial support by the doctoral fellowship program of the University of Rostock.

\begin{appendix}
\section{How the choice of the Bloch ansatz affects the coupling to external fields in Bloch-Hamiltonians and the current operator in $\vec k$-space for the SSH chain }\label{appendix:properblochansatz}

The simplest solid-state-like system that displays topological features is the SSH chain \cite{SSHPhysRevLett.42.1698,topinsshortcourse}. The tight-binding Hamiltonian \eqref{eq:tbH}
 for the SSH chain takes into account two orbitals per unit cell and only intracell ($\Delta\vec R=0$) and intercell hoppings ($\Delta\vec R= \vec a_1$) with amplitude $v$ and $w$, respectively,
\beq \hat H = \sum_m \bigl(v \ket{m,2} \bra{m,1} + w \ket{m+1,1}\bra{m,2} + \mathrm{h.c.} \bigr). \label{eq:SSH-Hamiltonian} \eeq 
Here, we simplified the notation, i.e., $\ket{\phi_{\vec R \alpha}} \to \ket{m,\alpha}$ where the cell index $m$ corresponds to $\vec R \to m \vec a_1 $ in 1D, and $\alpha=1,2$.
Just given a tight-binding Hamiltonian like \eqref{eq:SSH-Hamiltonian}, we have some freedom to make contact to actual position-space coordinates. The SSH model is usually thought of describing a dimerized chain where, starting from an equidistant atom distribution with a distance $a/2$, the atoms are shifted alternatingly by a small amount $\delta$ to the right and to the left, thus doubling the primitive cell to size $a$. As long as $\delta\ll a$ we can write $\vec r_{m2}-\vec r_{m1} \simeq \vec r_{m+1,1}-\vec r_{m2} \simeq a/2$ (where $\vec r_{m'\alpha'}-\vec r_{m\alpha}$ are the respective distances to hop). The Bloch ansatz \eqref{eq:Blochlikeansatz} then reads 
\beq \ket{k,\alpha} = \sum_n \eulere^{\imagi (n+(\alpha-1)/2)ak} \ket{n,\alpha}, \qquad \alpha=1,2, \label{eq:SSHBlochlikeansatz2}\eeq
i.e., the two sites within a unit cell are at positions $\tau_1=0$ and $\tau_2=a/2$.
Equation \eqref{eq:psink} becomes
\beq \ket{\pm, k} 
= \sum_{n\alpha} C^\alpha_{\pm}(k)\,\eulere^{\imagi (n+(\alpha-1)/2)ak}\, \ket{n,\alpha}, \label{eq:pmkansatz2} \eeq
and insertion into the position-space Hamiltonian \eqref{eq:SSH-Hamiltonian} yields
\beq  E_{\pm}(k) \mathbf{C}_\pm(k) = \mathbf{H}(k) \mathbf{C}_\pm(k), \label{eq:BlochforSSH2}\eeq
where 
\beq \mathbf{H}(k) = \twobytwo{0}{s^*(k)}{s(k)}{0}\label{eq:BlochHamilSSHalternative} \eeq 
with \beq s(k)=v\eulere^{-\imagi a k/2}+w^*\eulere^{\imagi a k/2}.  \eeq
The dispersion relation is
\beq E_\pm(k) = \pm \sqrt{s(k)s^*(k)}, \eeq 
possible normalized eigenvectors are
\beq \mathbf{C}_\pm(k) = \frac{1}{\sqrt{2}} \twovec{1}{\frac{E_\pm(k)}{s^*(k)}} .\label{eq:eigenvectorsSSH2}\eeq

It can be shown \cite{PhysRevB.51.4940} that the usual ``minimal substitution'' $\hat{\vec p} \to \hat{\vec p} + \vec A(\vec r, t)$ to couple an electron to an external driver described by a vector potential $\vec A(\vec r, t)$ in the continuous case amounts in tight binding to the replacement of the hopping elements
\begin{eqnarray*} \lefteqn{\ket{m',\alpha'}\bra{m,\alpha} } \\
 & \rightarrow & \eulere^{-\imagi (\vec r_{m'\alpha'}-\vec r_{m\alpha}) (\vec A_{m'\alpha'}(t) + \vec A_{m\alpha}(t))/2}\ket{m',\alpha'}\bra{m,\alpha} \label{eq:peierlssubsti}
 \end{eqnarray*}
where $ \vec A_{m\alpha}(t)$ is the vector potential at position $\vec r_{m\alpha}$. In dipole approximation, $\vec A_{m\alpha}(t)= \vec A(t)$ is independent of space such that the time-dependent Hamiltonian reads
\beq  \hat H(t) = \sum_m \bigl(v(t)  \ket{m,2} \bra{m,1} + w(t)   \ket{m+1,1}\bra{m,2} + \mathrm{h.c.} \bigr)  \label{eq:time-dep-SSH-Hamiltonian} \eeq
with
\beq v(t) = v \, \eulere^{-\imagi aA(t)/2}, \qquad w(t) =  w \, \eulere^{-\imagi aA(t)/2}. \eeq

We now try the ansatz \eqref{eq:pmkansatz2} but time-dependent,
\beq \ket{\Psi(k,t)} = \sum_{n\alpha} C^\alpha(k,t)\,\eulere^{\imagi (n+(\alpha-1)/2)ak}\, \ket{n,\alpha}, \label{eq:ansatzPsiktSSH} \eeq 
for the time-dependent Schr\"odinger equation
\beq \imagi\partial_t \ket{\Psi(t)} = \hat H(t)  \ket{\Psi(t)} \eeq
and find, indeed,
 \beq  \imagi \mathbf{\dot{C}}(k,t) = \mathbf{H}(k,t) \mathbf{C}(k,t) \label{eq:TDBlochforSSH}\eeq
with
\beq \mathbf{H}(k,t) = \twobytwo{0}{s^*(k,t)}{s(k,t)}{0}, \quad s(k,t)= s[k+A(t)].\label{eq:TDBlochSSH} \eeq
As expected, the laser is coupled by replacing $k\to k+A(t)$ in the Bloch-Hamiltonian. With the initial condition $\mathbf{C}(k,0) = \mathbf{C}_\pm(k)$ we can follow how a Bloch state $\mathbf{C}_\pm(k)$ evolves in the laser field.

The informed reader may notice that the Bloch-Hamiltonian \eqref{eq:BlochHamilSSHalternative} is not the one usually discussed in the literature when it comes to the topological properties of the SSH model \cite{SSHPhysRevLett.42.1698,topinsshortcourse}. The reason is that a simpler Bloch ansatz is often used, namely
\beq \overline{\ket{k,\alpha}} = \sum_n \eulere^{\imagi nak} \ket{n,\alpha}, \qquad \alpha=1,2 \label{eq:SSHBlochlikeansatz}\eeq
instead of \eqref{eq:SSHBlochlikeansatz2}. This Bloch ansatz leads to the same form of the Bloch-Hamiltonian \eqref{eq:BlochHamilSSHalternative} but with $s(k)$ replaced by
\beq  \bar s(k)= v   + \eulere^{\imagi ak}  w^*, \eeq
i.e., \beq \bar{\mathbf{H}}(k) = \twobytwo{0}{\bar s^*(k)}{\bar s(k)}{0}.\label{eq:BlochHamilSSH} \eeq 
Because
\beq s(k) = \eulere^{-\imagi ak/2} \bar s(k) \eeq
the eigenvalues do not change,
$ E_\pm(k) = \pm \sqrt{s(k)s^*(k)}= \pm \sqrt{\bar s(k)\bar s^*(k)}$
but the eigenvectors do,
\beq \bar{\mathbf{C}}_\pm(k) = \frac{1}{\sqrt{2}} \twovec{1}{\frac{E_\pm(k)}{\bar s^*(k)}} .\label{eq:eigenvectorsSSH}\eeq
In the time-dependent case, \eqref{eq:ansatzPsiktSSH} becomes
\beq \overline{\ket{\Psi(k,t)}} = \sum_{n\alpha} \bar{C}^\alpha(k,t)\,\eulere^{\imagi nak}\, \ket{n,\alpha}. \label{eq:ansatzPsiktSSHorig} \eeq

In order to calculate SSH spectra, we need to evaluate the (time-derivative of) the current or the velocity expectation value.
The current can be derived from the continuity equation using Gauss law and the Heisenberg equation of motion for the density operator \cite{topinsshortcourse}.
Because there is only nearest-neighbor hopping and only two sites per unit cell, the intracell current for the SSH chain is simple and reads 
 \beq \hat j_{m}(t) = -\imagi  \bigl( v^*(t)   \ket{m,1} \bra{m,2} -   v(t) \ket{m,2} \bra{m,1} \bigr).  \eeq
Here, the subscript $m$ indicates that this is the current between sites $1$ and $2$ within unit cell $m$. The intercell current through the boundary at $(m+1/2)a$ (i.e., to the right of cell $m$) reads 
\begin{align} 
\hat j_{m+1/2}(t) 
&= -\imagi \bigl( w^*(t)   \ket{m,2} \bra{m+1,1} \nonumber \\
& \qquad\qquad -   w(t) \ket{m+1,1} \bra{m,2} \bigr) . \label{eq:SSHintercurrentplus}\end{align}
Note that the current operators are time-dependent.
With the state \eqref{eq:ansatzPsiktSSH} follows for the expectation value of the total current 
 \begin{align} \langle \hat j(t)\rangle(t) &= 
 &=  -\frac{2}{a} \, \mathbf{C}^\dagger(k,t)  \big[ \partial_k \mathbf{H}(k) \big]_{k+A(t)} \mathbf{C}(k,t), \label{eq:SSHtotcurrent}\end{align}
which has the expected form $j=-env$ where $n$ is the particle density (in this case particles per length), $v=\dot x=\partial_k H$ is the velocity, and $-e$ is the electron charge ($=-1$ in a.u.).

Instead, with the Bloch ansatz \eqref{eq:SSHBlochlikeansatz} and \eqref{eq:ansatzPsiktSSHorig} one obtains for the intercell current
\begin{align}
\overline{\langle \hat j_{m+1/2}(t)\rangle}(t) 
&= -\frac{1}{a} \bar{\mathbf{C}}^\dagger(k,t) \big[ \partial_k \bar{\mathbf{H}}(k) \big]_{k+A(t)/2} \bar{\mathbf{C}}(k,t).
\end{align}
and for the intracell current
\begin{align}
\overline{\langle \hat j_{m}(t)\rangle}(t) 
&= \bar{\mathbf{C}}^\dagger(k,t) \bar{\mathbf{J}}_{\mathrm{intracell}}(k,t) \bar{\mathbf{C}}(k,t)\end{align}
where
\beq \bar{\mathbf{J}}_{\mathrm{intracell}}(k,t) = \twobytwo{0}{-\imagi v^*(t)}{\imagi v(t)}{0}. \eeq
We see that, employing the Bloch ansatz \eqref{eq:SSHBlochlikeansatz}, the intercell current is related to $\partial_k \bar{\mathbf{H}}(k)$. However, the replacement is $k\to k+A(t)/2$, and the intracell current is not captured by $\partial_k \bar{\mathbf{H}}(k)$. The conclusion thus is that one should use the Bloch ansatz \eqref{eq:SSHBlochlikeansatz2} because only with this ansatz the coupling to an external field is correctly implemented by the replacement $\vec k\to \vec k+\vec A(t)$ in the field-free Bloch-Hamiltonian, and the current calculated using the time-dependent Bloch-Hamiltonian agrees with the physically meaningful current derived from the continuity equation in position space.

In the book by Vanderbilt \cite{vanderbiltbook}, Sec.\ 2.2.3, the choice for the Bloch ansatz \eqref{eq:SSHBlochlikeansatz2} is referred to as ``convention I'' while the ansatz \eqref{eq:SSHBlochlikeansatz} (i.e., the omission of the intracell positions $\boldsymbol{\tau}_\alpha$ in \eqref{eq:Blochlikeansatz}) is ``convention II''. The choice of the convention not only has consequences for the consistent coupling of the Bloch-Hamiltonian to external fields but also for the calculation of topological invariants, as discussed in \cite{vanderbiltbook} as well. In the case of the SSH chain, a winding number can be defined that counts how many times the origin in $\vec d$-space is encircled while $k$ sweeps through the Brillouin zone from $-\pi/a$ to $\pi/a$. This picture works well with convention II, because (for $v,w\in\mathbb{R}$) we have $d_x=v+ w\cos(ak)$, $d_y=v+w\sin(ak)$, $d_z=0$ so that $\vec d(k)$ indeed describes a circle of radius $w$ centered at $\vec d=(v,0,0)$. It is then easy to see that for $w>v$, the origin is encircled once while for $v>w$ the origin lies outside the circle. In finite SSH chains, $w>v$ implies dangling sites at the chain's edges, leading to edge states. In that sense, the winding number---defined for the bulk---is a topological invariant, as it ``predicts'' the presence of edge states in the corresponding finite system. This is an example for the so-called ``bulk-boundary correspondence'' \cite{vanderbiltbook}.

Of course, from a pragmatic view-point a Bloch ansatz is just a mathematical trick to switch from position space to $\vec k$-space where the problem simplifies to an (in our case) $2\times 2$ Bloch-Hamiltonian (for each $\vec k$). One can choose either convention for the Bloch ansatz. While convention II might be more convenient for the discussion of topological properties, convention I is simpler and less error-prone for the coupling to external fields and when the calculation of physically meaningful currents $\sim [\boldsymbol{\nabla}_{\vec k} \mathbf{H}(\vec k)]_{\vec k +\vec A(t)}$ is required.

\section{Haldane model}\label{appendix:haldane}
\begin{figure}
\includegraphics{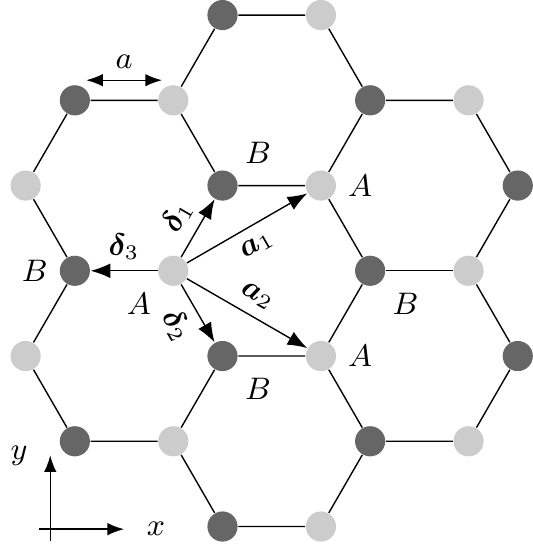}
\caption{Geometry of the hexagonal lattice used in the Haldane model. The unit cell consists of two sites \( A \) (light gray) and \( B \) (dark gray), each contributing one (tight-binding) orbital. The \( \boldsymbol{a}_i \), $i=1,2$, are lattice vectors and connect next-nearest neighbors, the \( \boldsymbol{\delta}_i \), $i=1,2,3$, connect nearest neighbors.}
\label{fig:haldane-model}
\end{figure}
The Haldane model describes a 2D hexagonal system with broken inversion and broken time-reversal symmetry such that it displays topological effects (without external magnetic field) \cite{Haldane88}.
The hexagonal lattice with two orbitals \(A\) and \(B\) per unit cell is shown in Fig. \ref{fig:haldane-model}.
The lattice vectors are
\begin{align}
	\boldsymbol{a}_1 &= \frac{a}{2} \twovec{3}{\sqrt{3}}
	, &
	\boldsymbol{a}_2 &= \frac{a}{2} \twovec{3}{-\sqrt{3}}
\end{align}
with the lattice constant \( a \). The nearest-neighbor vectors are
\begin{align}
	\boldsymbol{\delta}_1 &= \frac{a}{2} \twovec{1}{\sqrt{3}}
	, &
	\boldsymbol{\delta}_2 &= \frac{a}{2} \twovec{1}{-\sqrt{3}}
	, &
	\boldsymbol{\delta}_3 &= -a \twovec{1}{0}
	.
\end{align}
The tight-binding Hamiltonian is
\begin{widetext}
\begin{equation}
	\hat{H} =
	\sum_i M \left( \ket{i,A} \bra{i,A} - \ket{i,B} \bra{i,B} \right)
	+
	\sum_{<i,j>} t_1 \left( \ket{j,A} \bra{i,B} + \mathrm{h.c.} \right)
	+
	\sum_{\ll i,j \gg} \sum_{\alpha \in \{A,B\}} \left( t_2 \ket{j,\alpha} \bra{i,\alpha} + \mathrm{h.c.} \right) \label{eq:Haldanehamilposspace}
\end{equation}
\end{widetext}
with an alternating onsite potential \( M \) for the orbitals in the first sum (breaking inversion symmetry), nearest neighbor hopping with the real amplitude \( t_1 \) in the second sum, and complex next-nearest neighbor hopping with the amplitude \( t_2 \) (breaking time-reversal symmetry) in the third sum.
We choose the next-nearest neighbor hopping such that the term with $t_2$ describes counter-clockwise hopping while the term with $t_2^*$ describes clockwise hopping within one hexagon.

As discussed in Appendix \ref{appendix:properblochansatz}, convention I is simpler for the coupling to external fields and therefore intracell positions should be included in the ansatz
\begin{align}
	\ket{\pm, \boldsymbol{k}} = \sum_{mn} &\eulere^{\imagi \left(m \boldsymbol{a}_1 + n \boldsymbol{a}_2 \right)\cdot\boldsymbol{k}}\, \ket{m,n} \nonumber\\ 
	&\otimes \left(C^A_{\pm}(\boldsymbol{k})\,\ket{A} + C^B_{\pm}(\boldsymbol{k})\,\eulere^{\imagi \boldsymbol{\delta}_3\cdot\boldsymbol{k}}\,\ket{B}\right). \label{eq:BlochansatzHaldane}
\end{align}
After a straightforward calculation, and with the vectors
\begin{align}
	\boldsymbol{g}_1 &= \boldsymbol{a}_1 - \boldsymbol{a}_2 \, ,
	&
	\boldsymbol{g}_2 &= - \boldsymbol{a}_1 \, ,
	&
	\boldsymbol{g}_3 &= \boldsymbol{a}_2,
\end{align}
the Bloch-Hamiltonian \eqref{eq:haldane-bloch-hamiltonian} is obtained.
Introducing
\begin{equation}
\begin{split}
	\tau(\boldsymbol{k}) &= t_1 \sum_n \eulere^{\imagi \boldsymbol{\delta}_n \cdot \boldsymbol{k}},
	\\
	\kappa(\boldsymbol{k}) &= 2 \Re(t_2) \sum_n \cos(\boldsymbol{g}_n \cdot \boldsymbol{k}),
	\\
	\sigma(\boldsymbol{k}) &= M + 2 \Im(t_2) \sum_n \sin(\boldsymbol{g}_n \cdot \boldsymbol{k}),
\end{split}
\end{equation}
the Bloch-Hamiltonian can be written as
\begin{equation}
	\mathbf{H}(\boldsymbol{k}) = \twobytwo{\kappa + \sigma}{\tau}{\tau^*}{\kappa - \sigma},
\end{equation}
and the dispersion relation is
\begin{equation}
	E_{\pm}(\boldsymbol{k}) = \kappa \pm \sqrt{|\tau|^2 + \sigma^2} .
\end{equation}
Two possible sets of normalized eigenvectors are
\begin{equation}
\begin{split}
	\mathbf{C}_+^>(\boldsymbol{k}) &= \frac{1}{\sqrt{|\mathbf{C}^>|^2}} \twovec{\sigma + \sqrt{|\tau|^2 + \sigma^2}}{\tau^*}
	\\
	\mathbf{C}_-^>(\boldsymbol{k}) &= \frac{1}{\sqrt{|\mathbf{C}^>|^2}} \twovec{-\tau}{\sigma + \sqrt{|\tau|^2 + \sigma^2}}
\end{split}
\end{equation}
and
\begin{equation}
\begin{split}
	\mathbf{C}_+^<(\boldsymbol{k}) &= \frac{1}{\sqrt{|\mathbf{C}^<|^2}} \twovec{\tau}{-\sigma + \sqrt{|\tau|^2 + \sigma^2}}
	\\
	\mathbf{C}_-^<(\boldsymbol{k}) &= \frac{1}{\sqrt{|\mathbf{C}^<|^2}} \twovec{\sigma - \sqrt{|\tau|^2 + \sigma^2}}{\tau^*}
\end{split}
\end{equation}
with
\begin{equation}
\begin{split}
	|\mathbf{C}^>|^2 &= 2 \sqrt{|\tau|^2 + \sigma^2} \Bigl( \sigma + \sqrt{|\tau|^2 + \sigma^2} \Bigr)
	\\
	|\mathbf{C}^<|^2 &= 2 \sqrt{|\tau|^2 + \sigma^2} \Bigl( -\sigma + \sqrt{|\tau|^2 + \sigma^2} \Bigr)
\end{split}
\end{equation}
where \( \mathbf{C}_\pm^> \) is used for \( \sigma > 0 \) and \( \mathbf{C}_\pm^< \) for \( \sigma < 0 \).
This distinction based on the sign of \( \sigma \) is convenient to handle the limit \( |\tau| \rightarrow 0 \) numerically.

The coupling to an external driver described by a vector potential is performed with the Peierls substitution as in Appendix \ref{appendix:properblochansatz}.
Using the dipole approximation, the laser is again coupled by replacing \( \boldsymbol{k} \rightarrow \boldsymbol{k} + \boldsymbol{A}(t) \) in the Bloch-Hamiltonian.

\end{appendix}

	\bibliography{biblio.bib}

\end{document}